\documentclass[preprint,showpacs,preprintnumbers,amsmath,amssymb,aps]{revtex4}

\usepackage{graphicx}
\usepackage{amssymb}
\usepackage{amsmath}
\usepackage{bm}

\usepackage{color}

\begin{document}
\setcounter{page}{0}
\title[]{
Mechanisms of normal reflection at metal interfaces studied by Andreev-reflection spectroscopy}
\author{K. \surname{Gloos}}
\email{kgloos@utu.fi}
\thanks{Fax: +382-2333-5470}

\affiliation{Wihuri Physical Laboratory, Department of Physics   and Astronomy, University of Turku, FIN-20014 Turku, Finland}
\affiliation{Turku University Centre for Materials and Surfaces (MatSurf), FIN-20014 Turku, Finland}

\author{E. \surname{Tuuli}}
\affiliation{Wihuri Physical Laboratory, Department of Physics and Astronomy, University of Turku, FIN-20014 Turku, Finland}
\affiliation{The National Doctoral Programme in Nanoscience (NGS-NANO), FIN-40014 University of Jyv\"askyl\"a, Finland}

\date{\today}

\begin{abstract}
Andreev-reflection spectroscopy of elemental superconductors in contact with 
non-magnetic normal metals reveals that the strength of normal-reflection 
varies only slightly. 
This observation imposes strong constrictions on the three possible 
normal-reflection mechanisms: 
tunneling through a dielectric barrier, 
reflection due to the different electronic properties of the two electrodes,
and diffusive transport caused by elastic scattering in the contact region.
We discuss in detail the role played by Fermi-surface mismatch, 
represented by the different Fermi velocities on both sides of the contact 
interface. 
We find that it is at least not the dominant mechanism and possibly completely 
absent in the Andreev-reflection process.

\end{abstract}




\maketitle

\section{Introduction}

How do electrons cross a direct interface between a normal and a superconducting 
metal? 
At a tunnel junction the superconducting gap suppresses electron transport at 
small energies around the Fermi level. 
At a direct contact with no or only a very weak tunneling barrier, Andreev 
reflection transfers an additional electron for each incident one to form 
a Cooper pair in the superconductor. 
In an alternative picture the Cooper pair is composed of the incident
electron plus a newly created one by emitting a hole. 
Because of energy and momentum conservation this hole travels back
through the contact and into the normal conductor along almost the same path 
taken by the incident electron. 
Normal reflection as the natural counter part of Andreev reflection enhances 
the interface resistance, normal reflection of the second electron  or the retro-reflected hole  creates the characteristic double-minimum structure 
of the Andreev-reflection resistance spectra.
While the role of Andreev reflection in the transport process across the interface is understood \cite{Blonder1982}, the normal-reflection part is 
far from being settled. 

Normal reflection at a metal interface reduces its transmission 
coefficient $\tau = 1/(1+Z^2)$ to below unity. 
The dimensionless $Z$ parameter represents the strength of a $\delta$ -
function tunneling barrier that approximates a more realistic rectangular 
barrier of finite height $\Phi$ and width $w$ according to 
$Z_\delta = \Phi w / \hbar v_F$ for electrons with (average) Fermi 
velocity $v_F$ \cite{Blonder1982}.
Two other mechanisms contribute to normal-reflection. 
First, Fermi surface mismatch, again in terms of a $\delta$ - function barrier,
adds $Z_{FSM} = \left|{1-r}\right| /(2 \sqrt{r})$ where 
$r = v_{F1}/v_{F2}$ is the ratio of Fermi velocities of the two electrodes \cite{Blonder1983}. 
Second, diffusive transport through the contact region, conserves electron 
and hole energies in elastic scattering processes, but not their momentum. 
This allows partial backscattering of the incident electron as well
as the second electron or the retro-reflected hole.
A diffusive contact is thought to consist of a certain number of modes, 
each mode $i$ has its own transmission coefficient $\tau_i$. 
An ideal long diffusive junction has a distribution of transmission 
coefficients which sums up to a single $Z_{diff} \approx 0.55$
\cite{Artemenko1979,Mazin2001}.

Real contacts could have any combination of those three mechanisms.
Unless one of them dominates, separating the different contributions requires 
that they do not depend on each other, for example when the tunneling barrier 
sits at the end of the diffusive channel, or on one side of the interface and 
not on the other, with the total $Z \approx \sqrt{Z_{diff}^2+Z_{FSM}^2+Z_\delta^2}$.
Real contacts are probably more intricate than described by a 
$\delta$ - function barrier \cite{Blonder1983,Woods2004}. 
However, before invoking more involved and speculative modelling, one should 
try to explain the experimental data by starting with the most basic mechanisms
mentioned above.

Naidyuk {\it et al.} \cite{Naidyuk1996} as well as Naidyuk and
Yanson \cite{Naidyuk2005} have noticed that S (superconducting) - N (normal) 
point contacts, including those with unconventional heavy-fermion and 
high-temperature superconductors, often have $Z$ parameters between 0.4 
and 0.5, and that those contacts could be in the diffusive limit.
During the last couple of years we have found similar $Z \approx 0.5$ values
for many S - N combinations over a wide range of contact
resistances or lateral contact sizes that can not be explained by a 
dielectric barrier \cite{Tuuli2011,Gloos2012,Gloos2013}.
Since point-contacts between conventional metals and heavy-fermion compounds
with up to two orders of magnitude smaller $v_F$ also have 
$Z \approx 0.4 - 0.5$ \cite{Naidyuk2005}, Fermi-velocity mismatch 
does not seem to be a valid approach. 
That is why our initial interpretation of point-contact experiments with
superconducting niobium and conventional metals \cite{Tuuli2011}, which 
we will revise here, was based on the mismatch of Fermi momentum. 
Additionally, niobium provides such a wide margin of possible Fermi wave 
numbers from less than $4\,$nm$^{-1}$ up to $22\,$nm$^{-1}$ that one can 
almost freely pick a  suitable one \cite{Tuuli2011}.

We show here by comparing Andreev-reflection data of contacts between 
elemental superconductors and non-magnetic normal metals that Fermi-surface 
mismatch is not the dominant mechanism of normal reflection. 
Moreover, we suggest that even theoretically Fermi-surface mismatch 
does not affect normal reflection of the Andreev-reflected holes.

\section{Experiments and results}

We fabricated the point-contact interfaces using the shear (crossed wire) 
method by moving the two sample wires towards each other until they touch 
at one spot \cite{Chubov1982}.
Before the contact is set, the wires slide against each other and, thus, 
either remove or break up possible remains of an oxide layer, improving
the chance that the contacts are formed between relatively clean surfaces.
The normal conductors were silver (Ag), gold (Au), copper (Cu), 
palladium (Pd), and platinum (Pt), and the superconductors aluminum 
(Al, $T_c = 1.2\,$K), cadmium (Cd, 0.56 K), indium (In, 3.4 K), 
niobium (Nb, 9.2 K), tantalum (Ta, 4.4 K), tin (Sn, 3.7 K), titanium 
(Ti, 0.5 K), and zinc (Zn, 0.87 K).
The wires had a diameter of $0.25\,$mm except Al (0.5 mm), Cd (1.0 mm),
and In (1.5 mm).
Surface treatment did not noticeably affect the spectra with two exceptions 
that had otherwise enhanced $Z$ values or stronger normal reflection. 
The oxide layer of Nb was removed using fine abrasive paper, and that of Zn 
by dipping the wire in dilute HCl acid.
Before installing and cooling down, the wires were cleaned in an ethanol 
ultrasound bath.

\begin{figure}
  \includegraphics[width=8.5cm]{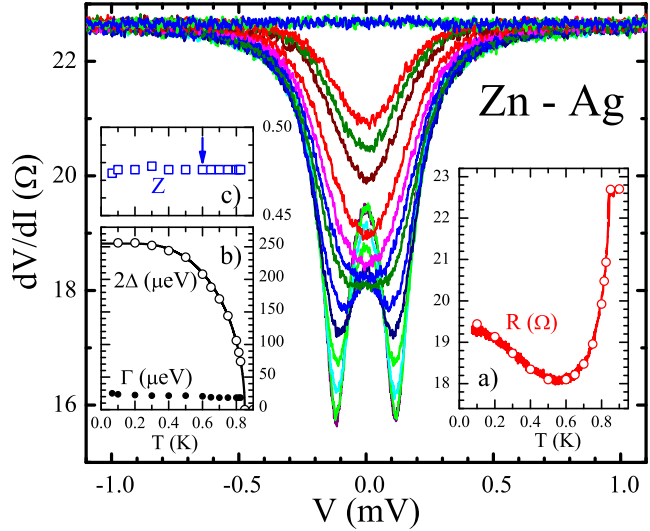}
  \caption{(Color online) a) Spectra of a Zn - Ag contact at discrete
  temperatures from 0.075 K to 0.90 K. 
  The insets show as function of temperature $T$: 
  a) Zero-bias resistance read off the spectra (open symbols) as 
  well as that recorded during the following cool down (solid line). 
  b) Energy gap $2\Delta$ and Dynes' parameter $\Gamma$. The solid line is 
  the theoretical BCS curve adjusted for $T_c = 0.85\,$K and 
  $2\Delta(T \rightarrow 0) = 255\,\mu$eV. 
  c) $Z$ parameter. 
  The arrow marks the temperature above which $Z$ was kept constant.}
  \label{spectra} 
  \end{figure}

The contacts were measured in the vacuum region of a dilution 
refrigerator.
A DC current $I$ with a small superposed AC component $dI$ ran
through the contact, and the voltage drop $V+dV$ across the 
contact was measured to obtain the $I(V)$ characteristics as well as the 
differential resistance spectrum $dV/dI(V)$.
In addition to S - N contacts we have also investigated S$_1$ - S$_2$ 
junctions between two different superconductors $S_1$ and $S_2$ that had 
critical temperatures $T_{c1} \gg T_{c2}$, for example Nb - Al or In - Zn, 
above $T_{c2}$ to drive superconductor $S_2$ normal while $S_1$ still 
remains superconducting.
Strong superconductors $S_1$ = In, Ta, and Nb needed temperatures well above 
$T_{c2}$ to suppress proximity-induced superconductivity in $S_2$.
The spectra were fitted using the modified BTK theory 
\cite{Plecenik1994,deWilde1996} that includes Dynes' lifetime parameter 
$\Gamma$ \cite{Dynes1978}.
This model contains only two other adjustable parameters, the energy gap 
$2\Delta$ and the $Z$ parameter.
The differential resistance at large bias voltages coincided with the 
normal contact resistance $R_N$.
Not all contacts revealed the typical Andreev-like spectra that could be 
easily analysed, but had additional anomalies. 
Find examples for the distribution between 'good' and 'bad' contacts
in Ref. \cite{Gloos2013}.

Figure \ref{spectra} shows representative spectra of a Zn - Ag contact 
as function of temperature together with the fit parameters. 
The energy gap $2\Delta(T)$ follows closely the theoretical BCS curve 
\cite{Thouless1960}, while $Z(T)$ and $\Gamma(T)$ barely depend on temperature.
The $Z$ parameter could be determined within $\Delta Z \approx \pm 0.01$
when the Andreev-reflection double-minimum structure is visible.
Near $T_c$ we kept $Z$ constant and adjusted only $2\Delta$ and $\Gamma$.

\begin{figure}
  \includegraphics[width=8.5cm]{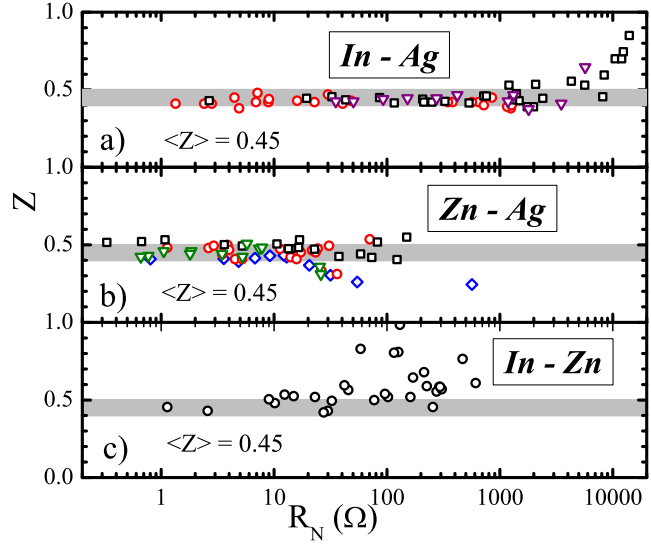}
  \caption{(Color online) a) $Z$ parameter of S - N contacts of a) In - Ag,
  b) Zn - Ag, and c) of S$_1$ - S$_2$ contacts In - Zn. 
  The In - Ag as well as the Zn - Ag contacts were measured $0.1\,$K, the 
  In - Zn contacts at $2.5\,$K to suppress   superconductivity in Zn. 
  Different symbols denote measurement series with different sample wires
  from the same batch.
  The thick solid lines represent an average $\langle Z \rangle = 0.45$ 
  with a $\pm 0.05$ bandwidth.}
  \label{Z-In-Zn} 
  \end{figure}

\begin{figure}
  \includegraphics[width=8.5cm]{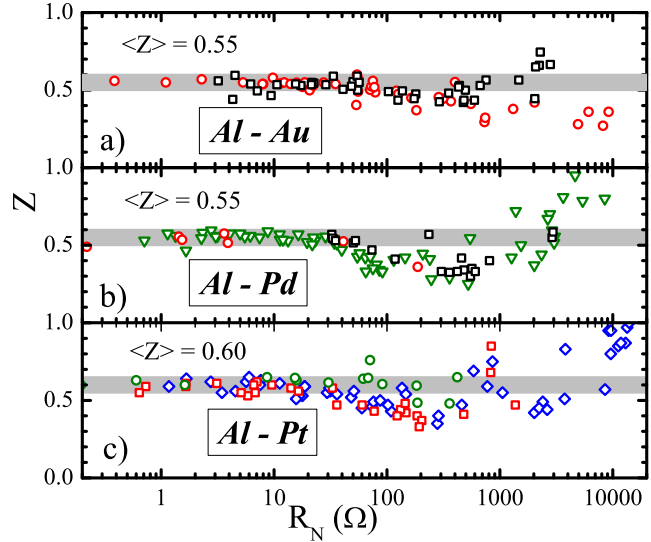}
  \caption{(Color online) a) $Z$ parameter of S - N contacts of a) Al - Au,
  b) Al - Pd, and c) Al - Pt. All contacts were measured at 0.1 K.
  The thick solid lines indicate the average $\langle Z \rangle$ with a 
  $\pm 0.05$ bandwidth.}
  \label{Z-Al-Au} 
  \end{figure}

\begin{figure}
  \includegraphics[width=8.5cm]{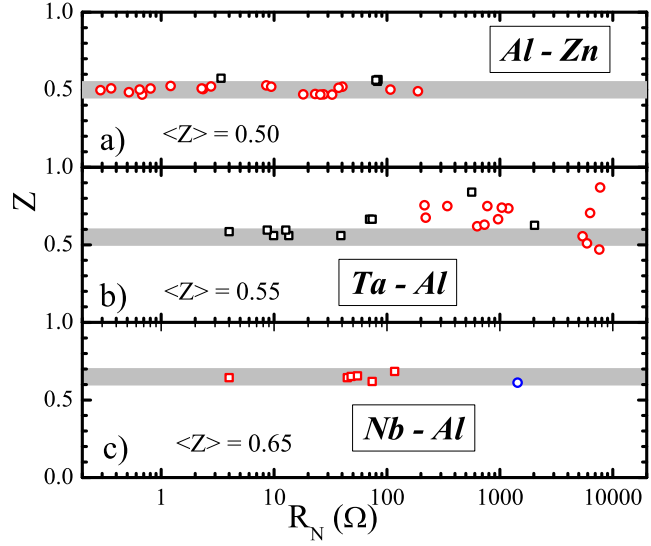}
  \caption{(Color online) a) $Z$ parameter of S$_1$ - S$_2$ contacts of 
  a) Al - Zn, b) Ta - Al, and c) Nb - Al. 
  The Al - Zn contacts were measured in the normal state of Zn around 0.85 K, 
  while the Ta - Al and Nb - Al contacts were measured at 2.5 K or higher to 
  suppress proximity-induced superconductivity in the otherwise normal Al.
  The thick solid lines indicate the average $\langle Z \rangle$ with a 
  $\pm 0.05$ bandwidth.}
  \label{Z-Al-S} 
  \end{figure}

\begin{figure}
  \includegraphics[width=8.5cm]{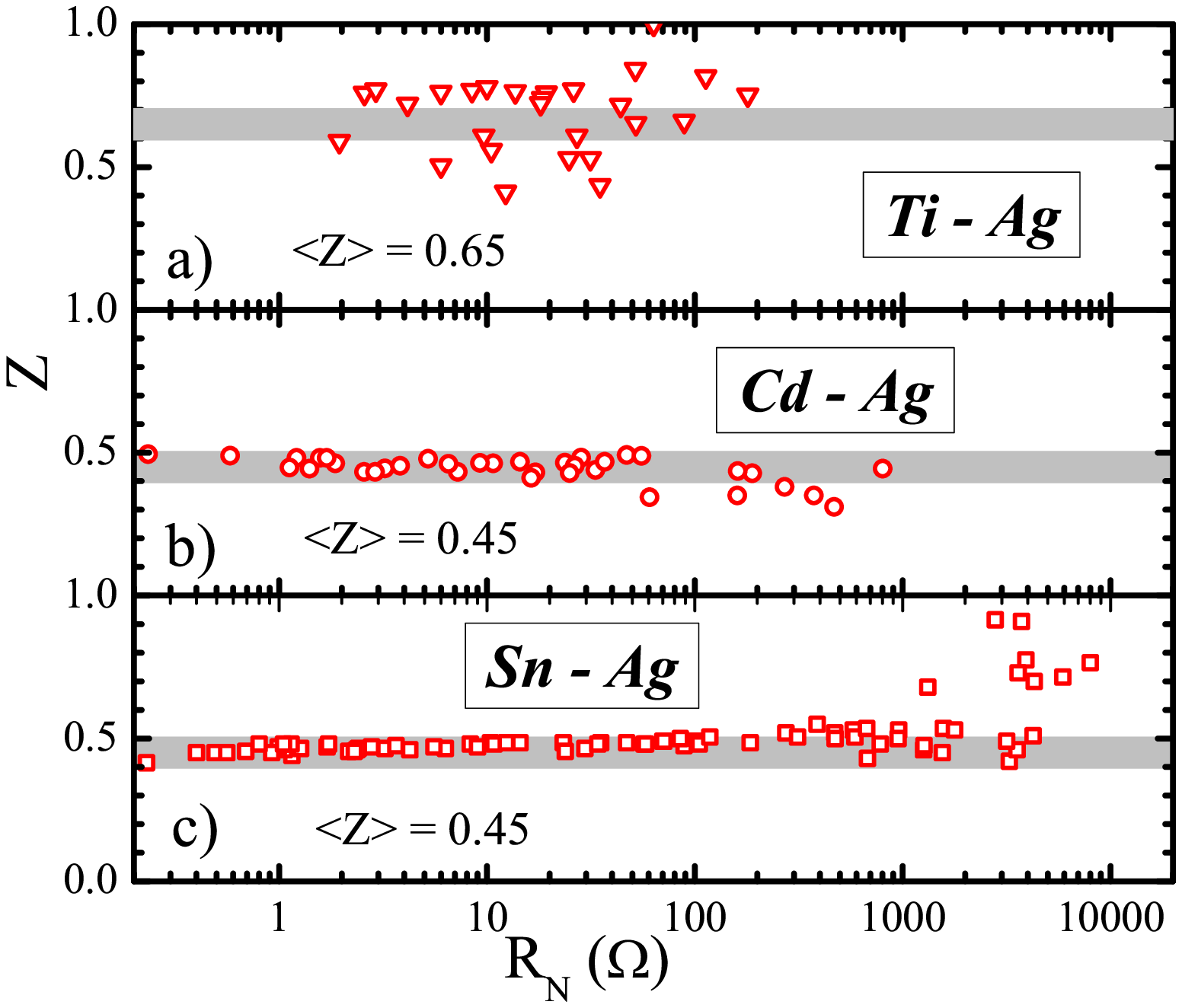}
  \caption{(Color online) a) $Z$ parameter of S - N contacts of a) Ti - Ag,
  b) Cd - Ag, and c) Sn - Ag, measured at 0.1 K.
  The thick solid lines indicate the average $\langle Z \rangle$ with a 
  $\pm 0.05$ bandwidth.
  The Ti - Ag data scatter over more than twice this width.}
  \label{Z-Ag-Ti} 
  \end{figure}

Figures \ref{Z-In-Zn} - \ref{Z-Ag-Ti} show the $Z$ parameters as function of
contact resistance $R_N$ for various S - N combinations.
Note that in many cases from small resistances of less than $1\,\Omega$ 
(contact diameter $\approx 30\,$nm) to above $1\,\text{k}\Omega$ 
(contact diameter $\approx 1\,$nm) the $Z$ parameter stays rather 
constant at around 0.5. 
At higher resistances above $1\,\text{k}\Omega$ the contacts typically had 
larger $Z$ values, probably because small interface areas are more susceptible 
to the surface quality. 
Also the cleaning mechanism of the shear contacts does not seem to work well 
with a soft counter electrode like In as shown in Figure \ref{Z-In-Zn} c)
where some of the In - Zn contacts had large $Z$ already at small $R_N$.
Some of the Al contacts in Figure \ref{Z-Al-Au} had $Z$ down to 0.3 in 
the $R_N \approx 100 - 1000\,\Omega$ range \cite{Gloos2012}.
Since this was accompanied by enhanced $2\Delta$  and $\Gamma$, we suspect 
that those small $Z$ values could be artefacts caused by inhomogeneous 
superconductivity in the contact region.

\section{Discussion}

Our main experimental results in Figures \ref{Z-In-Zn} - \ref{Z-Ag-Ti} 
confirm earlier observations that S - N contacts have rather similar $Z$ 
parameters around 0.5 
\cite{Naidyuk1996,Naidyuk2005,Tuuli2011,Gloos2012,Gloos2013}.
Explanations based on a dielectric tunneling barrier, Fermi-surface 
mismatch, or diffusive contacts, have to reproduce the very small variation 
of $Z$.

The question of a dielectric interface barrier can be resolved most easily. 
Such a barrier, or its remnants, could be present because we have prepared 
the contacts at ambient conditions. 
However, if tunneling would be the dominant mechanism one would expect 
a much larger variation of $Z$ not only from metal-to-metal combination 
but also from contact to contact \cite{Gloos2013}. 
The experimental data in Figures \ref{Z-In-Zn} - \ref{Z-Ag-Ti} demonstrate 
the opposite, therefore $Z_\delta \ll 0.5$ for contacts with 
$R_N \lesssim 1\,\text{k}\Omega$.
We can not exclude that tunneling plays a role for some of the 
In - Zn and Ti - Ag contacts which, nevertheless, agree with the lower 
bound of $Z \approx 0.4 - 0.5$.

The second reflection mechanism, Fermi-surface mismatch, can also not explain 
the small variation of the $Z$ parameter. 
Each time we make a new contact, the orientation of the crystallites that 
form the interface changes, and the $Z$ parameter should change accordingly. 
Since this is not observed, one should expect that $Z_{FSM} \ll 0.5$ 
\cite{Gloos2013}.

\begin{figure}
  \includegraphics[width=8.5cm]{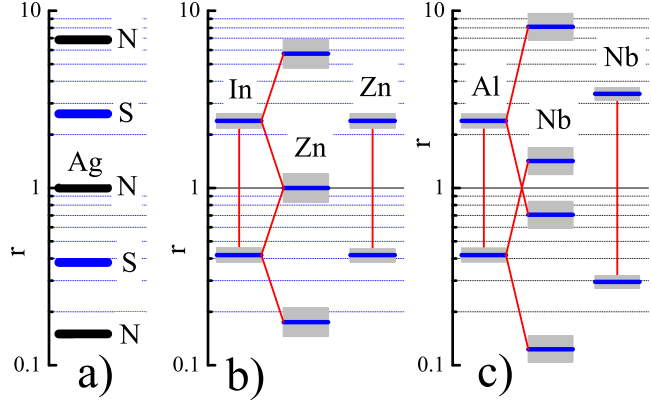}
  \caption{(Color online) 
  Fermi velocity $r$ (thick horizontal bars) normalized with respect to 
  that of Ag.
  a) Fermi-surface mismatch only causes all superconductors (S) to fall 
  into two categories with $r \approx 2.62$ or $r \approx 0.38$, assuming 
  an average $Z \approx 0.50$. 
  In turn all normal metals (N) either have $r \approx 6.85$, $1.00$, 
  or $0.16$. 
  b) Fermi velocities of In and Zn derived from In - Ag and Zn - Ag
  contacts as well as that of Zn calculated from In - Zn junctions.
  The shaded areas indicate error margins of $\Delta Z = \pm 0.05$.
  The two ways to derive $r$ of Zn clearly do not match.
  c) Likewise the Fermi velocities of Al and Nb are derived from
  Al - Ag and Nb - Ag contacts, then the velocity of Nb is calculated
  from Nb - Al contacts.}
  \label{triangles} 
  \end{figure}

A second argument against Fermi-surface mismatch comes from the magnitude 
of $Z$ for inter-related pairs of metals.
Take for example Zn - Ag and Zn - Cu contacts. 
They both have nearly the same $Z \approx 0.45$, that means the Fermi velocity 
of Ag and Cu would be either 2.39 or 1/2.39 = 0.42 times that of Zn, using
$Z = Z_{FSM} = \left|{1-r}\right| /(2 \sqrt{r})$.
From Ref. \cite{Ashcroft1976} we get Fermi-velocity ratios of 1.16 for the
Zn/Ag and and 1.32 for the Zn/Cu pair, respectively. 
These two sets of data do not agree well with each other.
The fact that Ag in contact with other superconductors and those 
superconductors in contact with other normal metals have rather similar
$Z \approx 0.5$ would imply the existence of just two groups of superconductors
that have $r \approx 2.62$ and $r \approx 0.38$ as well as three groups
of normal metals with $r \approx 2.62^2$, $r \approx 1.0 $, and 
$r \approx 0.38^2$ as indicated schematically in Figure \ref{triangles} a).
However, we have not found any S - N combination with $Z \gtrsim 1$ 
(velocity ratio $r \gtrsim 2.62^2$ or $r \lesssim 0.38^2$).
Therefore, all normal metals would have Fermi velocities near that of Ag, 
while all superconductors have either $r \approx 2.39$ or $r \approx 0.42$
with respect to Ag. 
This does not seem plausible.

Consider other combinations between one normal and two superconductors
like the Ag, In, and Zn triple. 
For In/Ag we get from Ref. \cite{Ashcroft1976} a velocity ratio of 1.25.
But now we can also measure normal Zn in contact with superconducting In.
They have the same $Z = 0.45$. 
Again use Ag as reference in Figure \ref{triangles} c). 
According to the first two experiments on Zn - Ag and In - Ag contacts, 
Zn and In should have Fermi velocities of either 2.39 or 0.42 times 
that of Ag. 
The third experiment with In - Zn contacts indicates that Zn should have a 
Fermi velocity of 2.39 or 0.42 times that of In, that is 
$0.42 \cdot 0.42 = 0.18$, $0.42 \cdot 2.39 = 1.00$, or $2.39 \cdot 2.39 = 5.72$ times that of Ag.
This contradicts the In - Ag data even if we allow for an uncertainty 
of $\Delta Z \approx \pm 0.05$. 
One can construct similar patterns also for other inter-related metal combinations, like for the Ag, Al, and Nb triple in Figure \ref{triangles} c). 
Thus Fermi surface mismatch in its common form \cite{Blonder1983} can not 
account for the observed $Z$ parameters.

How good is the free electron model for Andreev-reflection spectroscopy? 
It assumes a spherical Fermi surface and one, two, or three conduction 
electrons per atom. 
It describes reasonably well the alkali metals (which we do not use 
here) but not the transition metals like Nb and Ta or the nobel metals
Pd and Pd that have rather complex Fermi surfaces \cite{Choy2000}. 
Since electrons from different parts of the Fermi surface contribute to 
charge transport, it should be rather difficult to identify a specific 
Fermi velocity in those cases.

\begin{table}[position specifier]
\begin{center}
  \begin{tabular}{ c || c | c | c | c || c || c}
    \hline
    metal & $\rho$ & $\Theta_D$ &  $M$ & $\gamma$ & $v_{BG}$ & $v_F$ \\  
       & ($\mu\Omega\text{cm}$) & (K) & $\left( \frac{\text{g}}{\text{mol}} \right)$ & $\left( \frac{\text{mJ}}{\text{mol K}^2} \right)$ & (arb. units) &  ($\frac{\text{Mm}}{\text{s}}$) \\
    \hline \hline
    Ag & 1.59  & 221 & 108 & 0.646 & 7.35 & 1.39 \\ 
    \hline 
    Al & 2.65 & 390 & 27 & 1.35 & 4.47 & 2.02 \\ 
    \hline 
    Au & 2.21 & 178 & 197 & 0.69 & 5.55 & 1.38 \\ 
    \hline 
    Cd & 7.27 & 221 & 112 & 0.687 & 3.27 & 1.62 \\ 
    \hline 
    Cu & 1.68 & 310 & 64 & 0.695 & 6.41 & 1.57 \\ 
    \hline 
    In & 8.75 & 129 & 115 & 1.66 & 3.25 & 1.74 \\ 
    \hline 
    Nb & 14.5 & 260 & 93 & 7.8 & 0.64 & - \\ 
    \hline 
    Pd & 10.54 & 275 & 106 & 9.45 & 0.61 & - \\ 
    \hline 
    Pt & 10.6 & 225 & 195 & 6.54 & 0.65 & - \\ 
    \hline 
    Sn & 11 & 254 & 113 & 1.78 & 1.43 & - \\ 
    \hline 
    Ta & 13.15 & 225 & 181 & 5.87 & 0.64 & - \\ 
    \hline 
    Ti & 42 & 380 & 48 & 3.39 & 0.54 & 1.88 \\ 
    \hline 
    Zn & 5.96 & 237 & 65 & 0.64 & 4.59 & 1.82 \\ 
    \hline
  \end{tabular}
  \end{center}\caption{Bloch-Gr{\"u}neisen parameters of the various metals. 
  $\rho$ is the electrical resistivity at $T = 298\,$K, $\Theta_D$ the Debye temperature,
  $M$ the molar mass, and $\gamma$ the Sommerfeld constant of the electronic specific heat. 
  These data have been collected from \cite{knowledgedoor}.
  $v_{BG}$ is the Bloch-Gr{\"u}neisen derived velocity Eq. \ref{v_BG} and
   $v_F$ the free-electron Fermi velocity \cite{Ashcroft1976}.}
  \label{BG-parameters}
  \end{table}

Point-contact spectroscopy uses electrical transport properties as information source. 
Therefore it appears natural to extract an average Fermi velocity from the 
Bloch-Gr{\"u}neisen law for the temperature dependence of the electrical 
resistivity \cite{Ziman1960}. 
At high temperatures the resistivity $\rho(T)$ varies linearly with 
temperature, and does not depend on impurities.
The proportionality factor contains the Fermi velocity, the size of 
the Fermi surface of the conduction electrons, the surface area of the 
Debye sphere, and the strength of the electron-phonon interaction. 
One can write \cite{Varshney2006,Choudhary2007}
\begin{equation}
  \rho(T) \propto \frac{T}{\gamma v_F^2 \Theta_D^2 ML}
  \end{equation}
where $\gamma$ is the Sommerfeld constant of the electronic specific 
heat, $\Theta_D$ the Debye temperature, $M$ the molar mass, and 
$L$ a length scale that depends on the electron-phonon interaction.
Although the Bloch-Gr{\"u}neisen law describes relative changes of 
the resistivity quite well \cite{Ziman1960}, it is not commonly used 
to extract absolute values.
However, for our purposes we need only ratios of Fermi velocities, and
can get rid of the less well known $L$'s by assuming they are roughly 
the same for all metals, defining
\begin{equation}
  v_{BG} = \sqrt{  \frac{T}{\rho(T) \gamma \Theta_D^2 M}}
  \label{v_BG}
  \end{equation}
to use instead of the Fermi velocity.
Table \ref{BG-parameters} summarizes the parameters which allow to calculate 
$v_{BG}$ and thus the velocity ratios $r_{BG} = v_{BG1}/v_{BG2}$ for all 
investigated metal combinations.
We refrain from estimating error margins or how well $v_{BG}$ maps
the true $v_F$ and consider it as a number that characterizes 
the Fermi surface.
Note that $v_{BG}$ does not vary much for those metals which have a 
free-electron $v_F$ given in Ref. \cite{Ashcroft1976} except for Ti.
$v_{BG}$ is large for metals with a nearly spherical Fermi surface,
but small for all others.
A rather small $v_F$ of Ti has been noted by Zhang Dianlin {\it et al.} \cite{ZhangDianlin2005} and Hao Zhu {\it et al.} \cite{HaoZhu2013}. 
Also Nb is claimed to have a small Fermi velocity \cite{Kerchner1981}.

\begin{figure}
  \includegraphics[width=8.5cm]{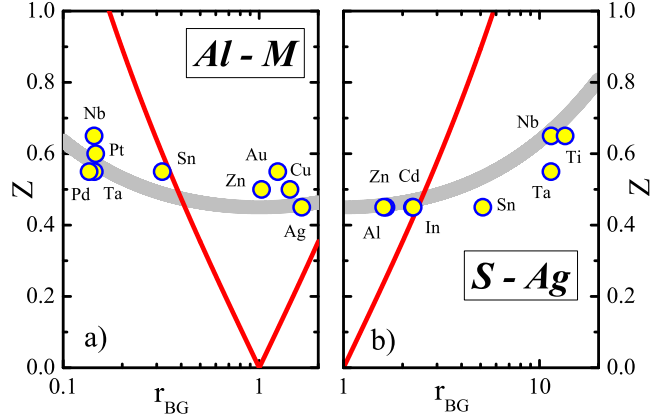}
  \caption{(Color online) The average $Z$ parameters of contacts between
  a) Al and other metals M and b) superconductors S and Ag as function of 
  their velocity ratio $r_{BG} = v_{BG,M}/v_{BG,Al}$ and 
  $v_{BG,S}/v_{BG,Ag}$, respectively. 
  The thin solid line describes the expected $Z$ due to Fermi surface 
  mismatch \cite{Blonder1983}, the thick solid is a guide to the eye 
  using   $Z^2 = 0.45^2 + 0.1 \cdot (1-r_{BG})^2/4r_{BG}$.}
  \label{Z_of_r_BG} 
  \end{figure}

Figure \ref{Z_of_r_BG} compares theory and experiment. 
It demonstrates that the $Z$ parameter depends only weakly on the velocity 
ratio, and that there is a huge background.
Therefore another mechanism must be responsible for the main contribution 
to $Z$.
Assume an independent one represented by $Z_0$ and write
\begin{equation}
  Z^2 \approx Z_0^2 + \alpha \cdot \frac{(1-r_{BG})^2}{4r_{BG}}
  \end{equation}
with a weight factor $\alpha$.
The data in Figure \ref{Z_of_r_BG} indicate $\alpha \approx 0.1$ so that
one can not reliably separate Fermi-surface mismatch even if it would exist.

In summary, all three arguments, {\it i)} the little variation of $Z$ 
from contact to contact, {\it ii)} the non-matching inter-related pairs 
of S - N junctions, as well as {\it iii)} the small variation of $Z$ with 
the velocity ratio $r_{BG}$ take the same line that Fermi-surface mismatch 
is not the dominant normal reflection mechanism.
Note that arguments {\it i)} and {\it ii)} apply not only to the velocity
mismatch but also to the case of the momentum mismatch.

This leaves the third reflection mechanism. Diffusive contacts have 
$Z \approx 0.55$ in the ideal case of a long channel 
\cite{Artemenko1979,Mazin2001}.
Thus diffusion  would conveniently explain our results, especially the 
large background $Z_0$.
However, a more realistic short instead of long diffusive channel or a 
strongly disordered interface should lead to different $Z$ values. 
We will investigate this possibility elsewhere.

Why is Fermi-surface mismatch absent, or at least strongle suppressed?
Electrons as well as Andreev-reflected holes travel through a 
dielectric tunneling barrier with a certain probability. 
Those not transmitted are reflected and enhance the contact resistance.
The retro-reflected holes that move towards the dielectric barrier can 
tunnel through with the same probability as the incident electrons.
This produces the typical Andreev reflection double-minimum structure
of the resistance spectra from which one can extract $Z$. 
The same applies to diffusive processes.
However, Fermi-surface mismatch works differently because it allows electrons
that have corresponding states on the other side to cross the interface 
and bars all others.
That means the transmission probability is either 1 or 0.
A retro-reflected hole is not affected again by Fermi-surface 
mismatch because it has already the right properties to find a corresponding 
state on the other side. 
Thus Fermi-surface mismatch affects the absolute value of the contact 
resistance but not the shape of the Andreev-reflection spectra.
Fermi-surface mismatch might play some role in the case of spin-polarized normal 
metals because there the two spin species have different Fermi surfaces,
and the retro-reflected holes flying back to the contact can face conditions 
that differ from that of the incident electrons.

\section{Conclusion}
We have found that Fermi-surface mismatch barely affects the 
normal-reflection part of Andreev reflection. 
This makes Andreev-reflection spectroscopy unsuitable for measuring
relative Fermi velocities of the electrodes, and 
leaves diffusive transport through the contact, that is elastic 
scattering at or near the contact interface, as the most probable 
mechanism to explain the usually observed double-minimum Andreev-reflection
anomalies.

\begin{acknowledgements}
We thank the Jenny and Antti Wihuri Foundation for financial support.
\end{acknowledgements}


\end{document}